\documentclass{article}
\usepackage[a4paper, total={6in, 8in}]{geometry}
\usepackage[ruled,vlined]{algorithm2e}
\usepackage{authblk}
\usepackage{braket}
\usepackage{mathtools}
\usepackage{caption}
\usepackage{subcaption}
\usepackage[
colorlinks=true,
urlcolor=blue,
citecolor=blue,
linkcolor=blue,
hyperfootnotes=false]{hyperref}
\usepackage{cleveref}
\usepackage{listings}
\usepackage{amsfonts}
\usepackage{url}

\SetKwProg{Fn}{Function}{}{}

\SetKwComment{Comment}{$\triangleright$\ }{}

\DeclareFixedFont{\ttb}{T1}{txtt}{bx}{n}{9} %
\DeclareFixedFont{\ttm}{T1}{txtt}{m}{n}{9}  %

\usepackage{color}
\definecolor{deepblue}{rgb}{0,0,0.5}
\definecolor{deepred}{rgb}{0.6,0,0}
\definecolor{deepgreen}{rgb}{0,0.5,0}

\newcommand\pythonstyle{\lstset{
language=Python,
basicstyle=\ttm,
morekeywords={self},              %
keywordstyle=\ttb\color{deepblue},
emph={MyClass,__init__},          %
emphstyle=\ttb\color{deepred},    %
stringstyle=\color{deepgreen},
showstringspaces=false
}}

\lstnewenvironment{python}[1][]
{
\pythonstyle
\lstset{#1}
}
{}

\newcommand\pythoninline[1]{{\pythonstyle\lstinline!#1!}}

\begin{document}

	\title{PyMatching: A Python package for decoding quantum codes with minimum-weight perfect matching}

	\author[1]{Oscar Higgott\thanks{oscar.higgott.18@ucl.ac.uk}}
	\affil[1]{Department of Physics \& Astronomy, University College London, WC1E 6BT London, United Kingdom}
	\date{\today}
	
	\maketitle
	
	\begin{abstract}
	This paper introduces PyMatching, a fast open-source Python package for decoding quantum error-correcting codes with the minimum-weight perfect matching (MWPM) algorithm.
	PyMatching includes the standard MWPM decoder as well as a variant, which we call \textit{local matching}, that restricts each syndrome defect to be matched to another defect within a local neighbourhood. 
	The decoding performance of local matching is almost identical to that of the standard MWPM decoder in practice, while reducing the computational complexity approximately quadratically.
	We benchmark the performance of PyMatching, showing that local matching is several orders of magnitude faster than implementations of the full MWPM algorithm using NetworkX or Blossom V for problem sizes typically considered in error correction simulations.
	PyMatching and its dependencies are open-source, and it can be used to decode any quantum code for which syndrome defects come in pairs using a simple Python interface.
	PyMatching supports the use of weighted edges, hook errors, boundaries and measurement errors, enabling fast decoding and simulation of fault-tolerant quantum computing.
	\end{abstract}

\section{Introduction}

Quantum error correcting codes will be necessary to protect large-scale quantum computers from noise.
An important piece of software required for using any quantum error correcting code is a \textit{decoder}, which takes as input the outcome of a set of check operator measurements (the syndrome) and attempts to find a correction operator that removes any error that may have occurred.

One approach to solving the decoding problem is to attempt \textit{minimum-weight} decoding, which finds the smallest error consistent with the syndrome.
For general quantum codes, there is no known solution for the minimum-weight decoding problem since this problem is known to be NP-complete~\cite{hsieh2011np,kuo2012hardness,kuo2020hardnesses,iyer2015hardness}.
However, for a wide class of quantum codes, the minimum-weight decoding problem (for either $X$ or $Z$ errors) can be solved efficiently with the help of Edmond's blossom algorithm~\cite{edmonds1965paths} for finding a minimum-weight perfect matching (MWPM) in a graph.
Quantum error correcting codes that can be decoded with MWPM include toric and surface codes~\cite{dennis2002topological}, the subsystem surface code~\cite{bravyi2012subsystem}, 2D hyperbolic~\cite{breuckmann2016constructions} and subsystem hyperbolic codes~\cite{higgott2020subsystem}, 3D toric and surface codes (for $X$ errors), the XZZX surface code~\cite{bonilla2020xzzx} and some compass codes~\cite{li20192d}, including the heavy hexagon code~\cite{chamberland2020topological}.
MWPM can also be used as a subroutine for single-shot decoding of the 3D toric code~\cite{quintavalle2020single} and the gauge color code~\cite{brown2016fault}, as well as for decoding the color code~\cite{kubica2019efficient}, fracton topological codes~\cite{brown2020parallelized}, the Fibonacci code~\cite{nixon2021correcting}, and the repetition code with noisy syndrome measurements.

A fast implementation of MWPM is often essential for simulations of quantum error correction, since estimating logical error rates accurately can require using a large number of Monte Carlo trials.
Efficient implementations of the blossom algorithm are provided by the Blossom V~\cite{kolmogorov2009blossom} and Lemon~\cite{dezsHo2011lemon} C++ libraries.
However, the blossom algorithm is only a subroutine in the MWPM decoder, as path finding algorithms are also required.
Implementations of a MWPM decoder include Autotune~\cite{fowler2012topological} and qecsim~\cite{qecsim}, both of which are tailored to specific variants of the toric and surface codes.

This paper introduces PyMatching, a fast, open-source Python package for decoding quantum error correcting codes with the MWPM decoder.
PyMatching includes the full MWPM decoder as well as a variant of MWPM, called local matching, that has significantly improved computational complexity compared to exact matching while retaining approximately the same decoding performance in practice.
While the core algorithms are implemented in C++, the functionality is available via a simple Python interface.
Furthermore, PyMatching can be used to decode any quantum code for which the MWPM decoder can be applied, rather than being tailored to specific quantum codes or noise models.
The source code for PyMatching can be found on Github\footnote{\url{https://github.com/oscarhiggott/PyMatching}}, and comprehensive documentation is also available\footnote{\url{https://pymatching.readthedocs.io/}}.
PyMatching is distributed under the open-source Apache 2.0 software license.

\section{Background}

Elements of the Paui group $\mathcal{P}_n$ are $n$-fold tensor products of Pauli operators in $\langle X,Y,Z\rangle$.
A stabiliser code is defined by a stabiliser group $\mathcal{S}$ which is an abelian subgroup of $\mathcal{P}_n$ that does not contain $-I$.
The codespace $\mathcal{T}(\mathcal{S})$ of a stabiliser code is the joint $+1$-eigenspace of elements of $\mathcal{S}$:
\begin{equation}
\mathcal{T}(\mathcal{S})\coloneqq\{\ket{\psi}~\mathrm{s.t.}~S\ket{\psi}=\ket{\psi}~\forall S\in \mathcal{S}\}.
\end{equation}

When using a stabiliser code for error correction, a set $\{S_1,S_2,\ldots,S_r\}$ of generators of $\mathcal{S}$, called check operators, are measured in order to obtain a syndrome.
Given a Pauli error $E\in\mathcal{P}_n$, its syndrome $\sigma(E)$ is a binary vector for which the $i$th element $\sigma(E)_i$ is 0 if $ES_i=S_iE$ and $\sigma(E)_i=1$ if $ES_i=-S_iE$.
The centraliser $C(\mathcal{S})$ of $\mathcal{S}$ in $\mathcal{P}_n$ is the set of Pauli operators that commute with every element of $\mathcal{S}$, and so an undetectable logical error is an element of $C(\mathcal{S})\setminus \mathcal{S}$.
The weight $|P|$ of a Pauli operator $P\in\mathcal{P}_n$ is the number of qubits on which it acts non-trivially, and the minimum distance of a stabiliser code is the minimum weight of any operator in $C(\mathcal{S})\setminus \mathcal{S}$.
A CSS stabiliser code has a stabiliser group that admits a set of generators $\{S_1,S_2,\ldots,S_r\}$ each satisfying $S_i\in\{I,X\}^n\cup \{I,Z\}^n$.

Given an error $E\in\mathcal{P}_n$, a decoder uses its syndrome $\sigma(E)$ to choose a correction operator $R\in\mathcal{P}_n$ to apply to the corrupted state $E\ket{\phi}$.
The decoder succeeds if $RE\in\mathcal{S}$ and fails otherwise.
If the correction $R$ is consistent with the syndrome then $RE\in C(\mathcal{S})$, and a logical error has occurred if $RE\in C(\mathcal{S})\setminus \mathcal{S}$.
The decoder therefore succeeds if it returns any element of the coset $[E]\coloneqq \{ES\quad\forall S\in\mathcal{S}\}$.
We can split the set of all possible errors consistent with the syndrome $\{EM: M\in C(\mathcal{S})\}$ into a disjoint union of cosets of the form $[E\bar{P}]$ for $\bar{P}\in C(\mathcal{S})\setminus \mathcal{S}$.
A \textit{maximum likelihood} decoder then finds the most probable coset, given an error model that assigns a probability $\pi(E)$ to each Pauli error $E\in\mathcal{P}_n$.
While optimal, maximum likelihood decoding is typically not efficient to implement (and is known to be \#P-Complete in general~\cite{iyer2015hardness}), although it can be well approximated for the surface code using the BSV decoder~\cite{bravyi2014efficient}.

A minimum weight decoder instead finds the minimum weight error consistent with the syndrome.
While the performance of minimum weight decoding is less good than maximum likelihood decoding, the minimum-weight perfect matching (MWPM) decoder can be used to solve the minimum weight decoding problem efficiently (for either $Z$ or $X$ errors) for some important families of quantum codes.

\section{Minimum-Weight Perfect Matching Decoder}

\begin{figure}
     \centering
     \begin{subfigure}[b]{0.32\textwidth}
         \centering
         \includegraphics[trim={3cm 0 1.5cm 0},width=\textwidth]{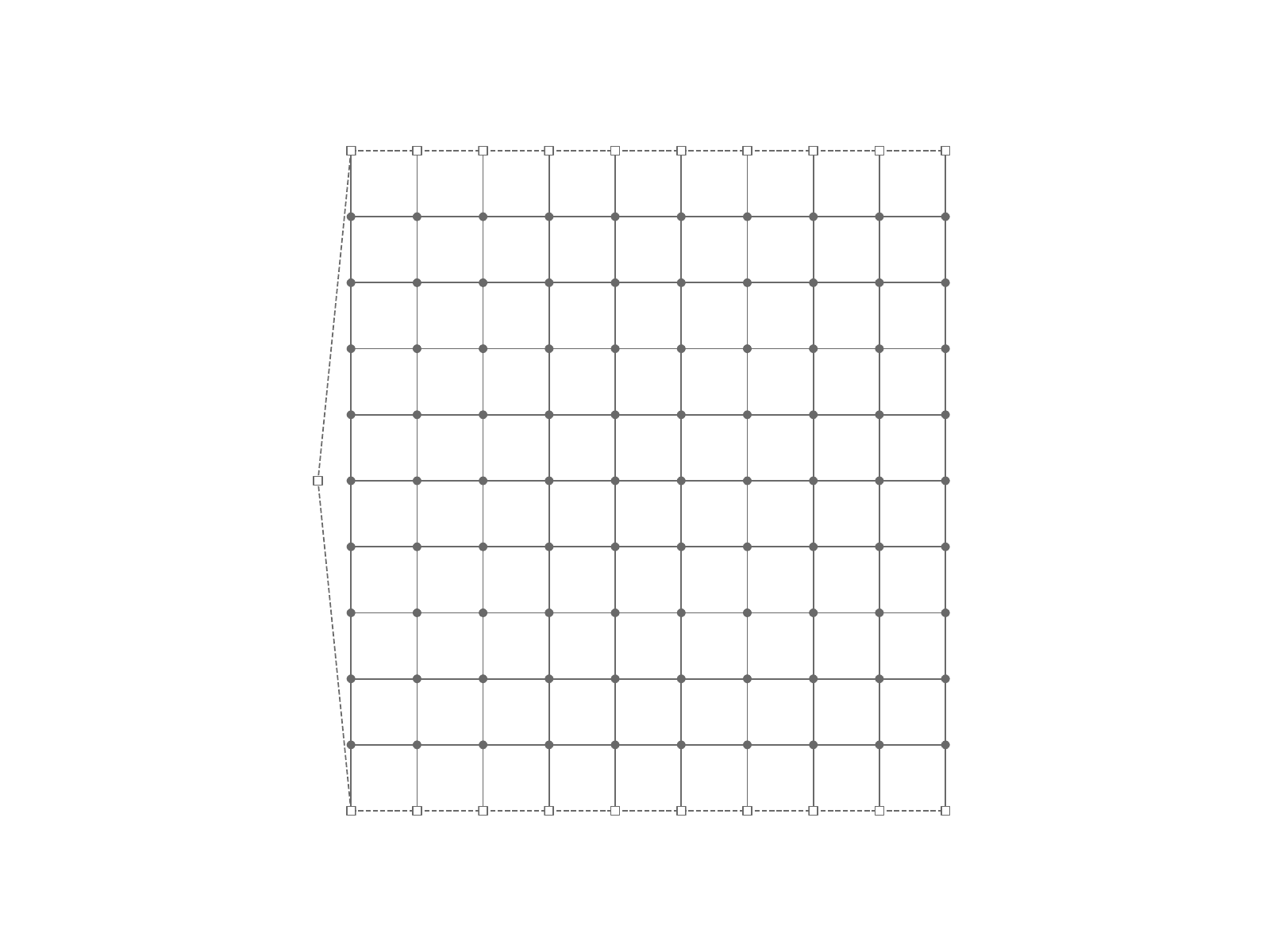}
         \caption{Matching graph}
         \label{fig:matching_graph}
     \end{subfigure}
     \hfill
     \begin{subfigure}[b]{0.32\textwidth}
         \centering
         \includegraphics[trim={3cm 0 1.5cm 0},width=\textwidth]{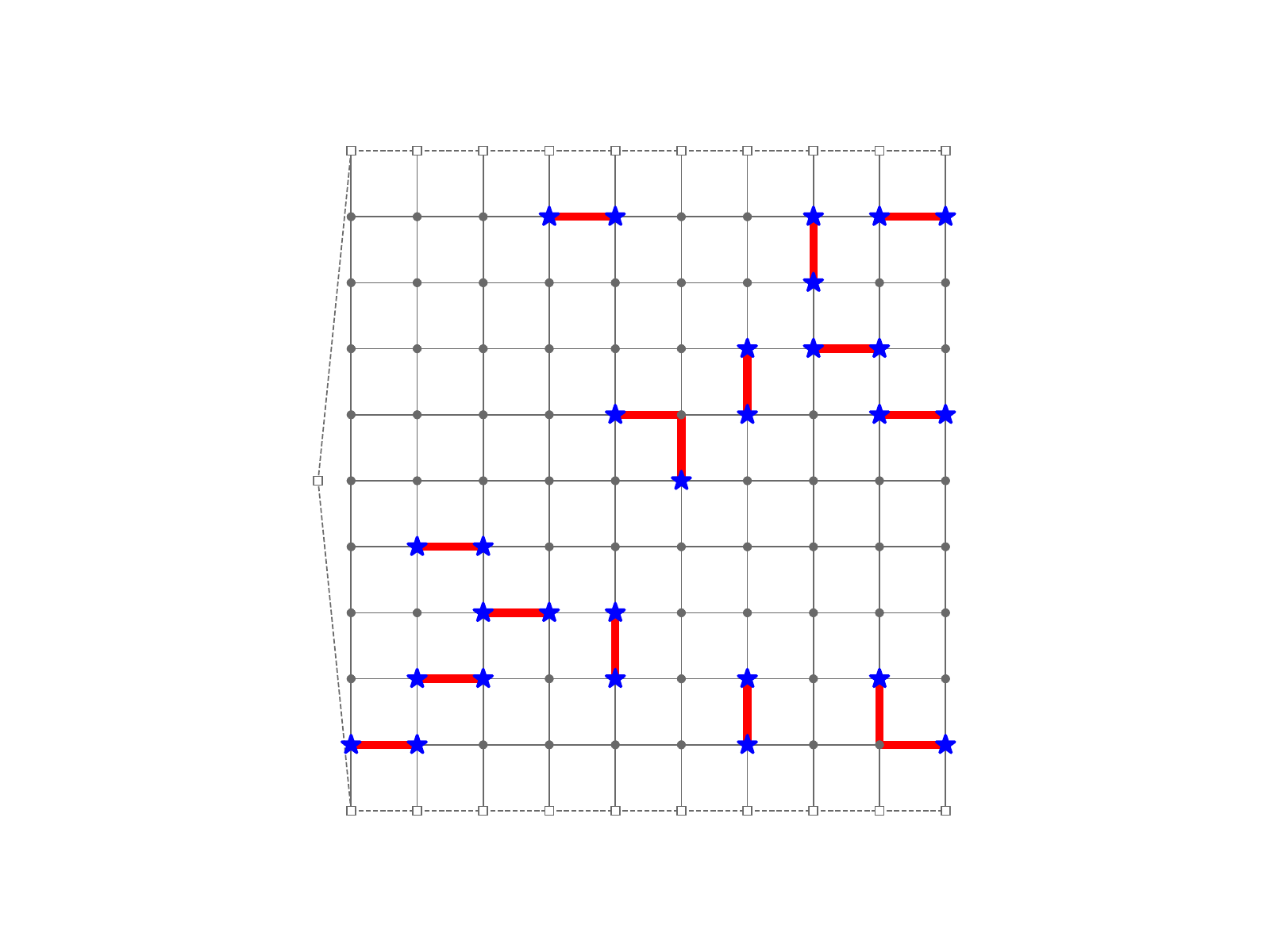}
         \caption{Error}
         \label{fig:error_graph}
     \end{subfigure}
     \hfill
     \begin{subfigure}[b]{0.32\textwidth}
         \centering
         \includegraphics[trim={3cm 0 1.5cm 0},width=\textwidth]{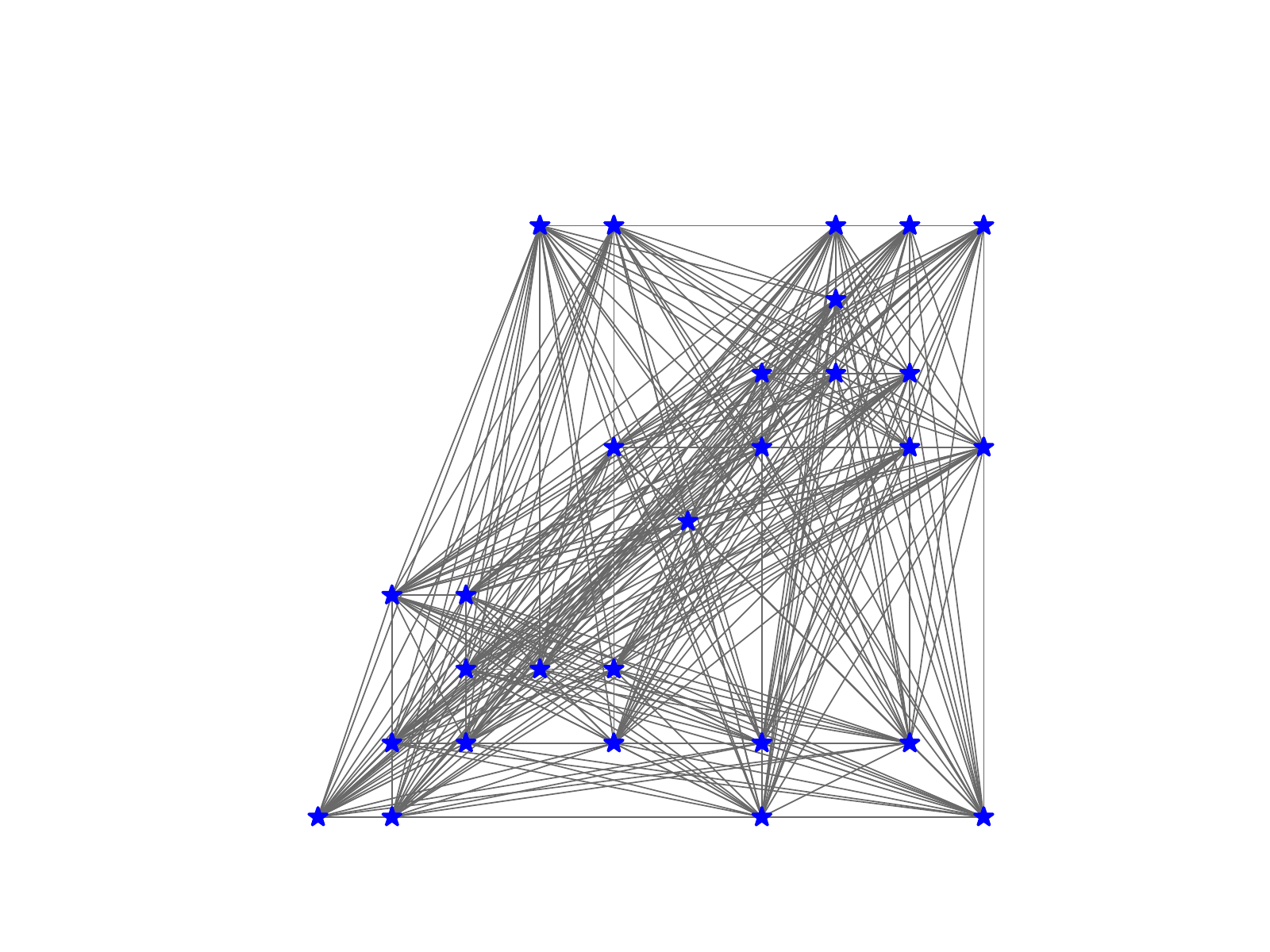}
         \caption{Syndrome graph}
         \label{fig:syndrome_graph}
     \end{subfigure}
     \hfill
     \begin{subfigure}[b]{0.32\textwidth}
         \centering
         \includegraphics[trim={3cm 0 1.5cm 0},width=\textwidth]{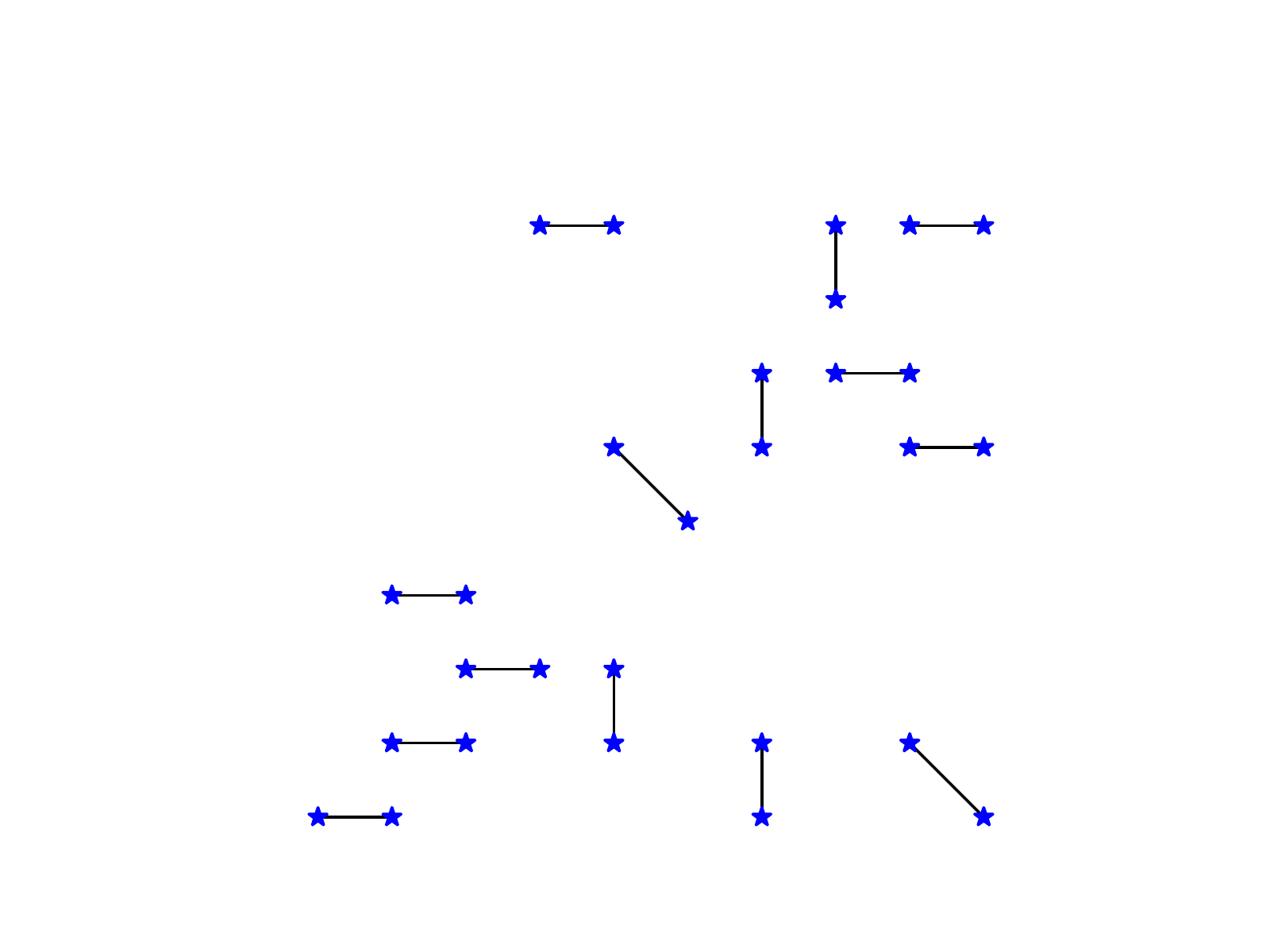}
         \caption{Minimum-weight perfect matching}
         \label{fig:mwpm_graph}
     \end{subfigure}
     \quad\quad\quad
     \begin{subfigure}[b]{0.32\textwidth}
         \centering
         \includegraphics[trim={3cm 0 1.5cm 0},width=\textwidth]{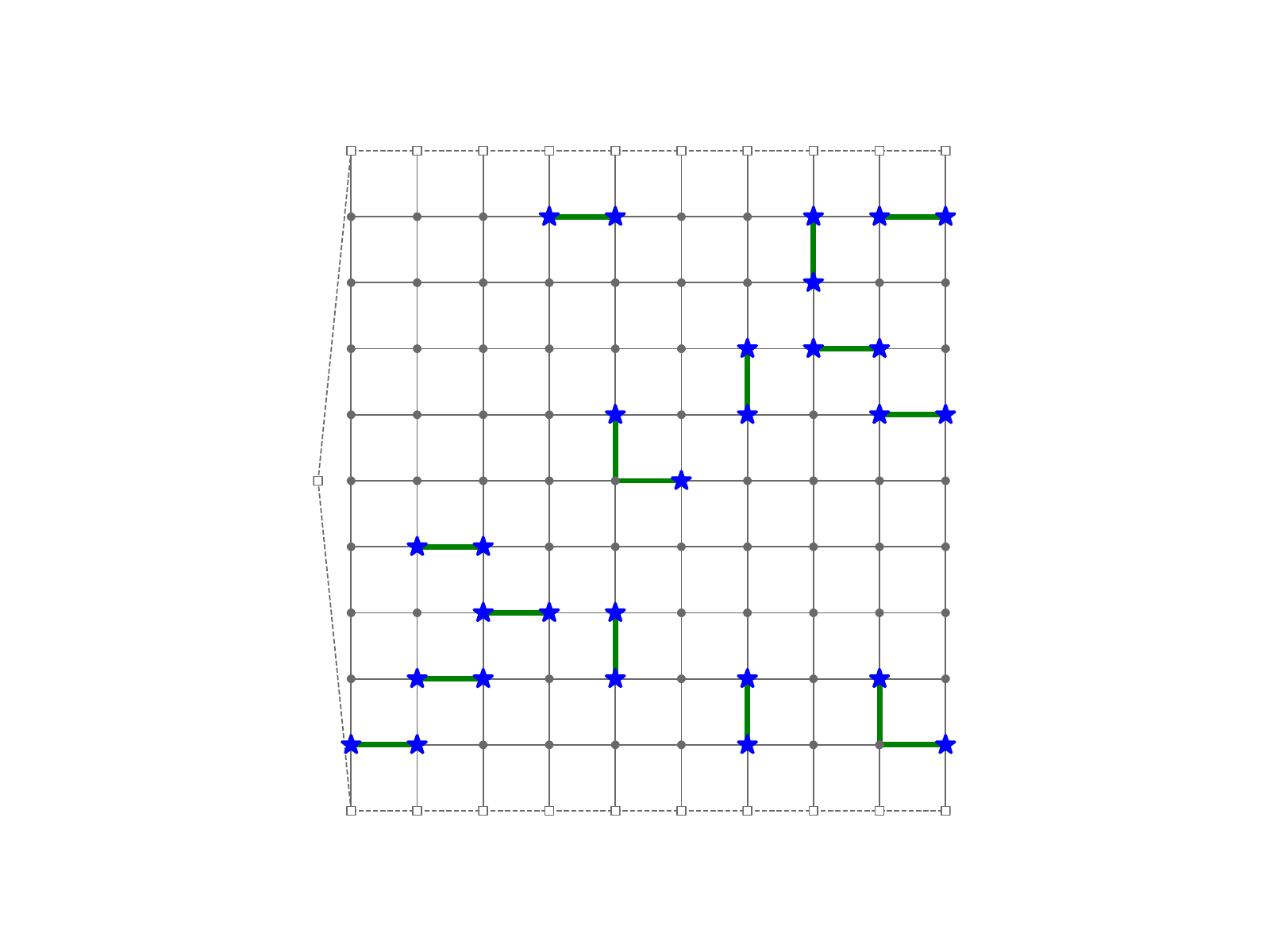}
         \caption{Correction}
         \label{fig:correction_graph}
     \end{subfigure}
        \caption{Stages of the minimum-weight perfect matching decoder for a distance 10 surface code.}
        \label{fig:mwpm_decoder}
\end{figure}

We will now consider the problem of decoding a Pauli $Z$ error $E\in\{I,Z\}^n$ for a CSS stabiliser code.
Note that we can decode Pauli $X$ errors with the same method, and MWPM can also be easily adapted to decode certain non-CSS codes (such as the XZZX surface code~\cite{bonilla2020xzzx}).
We denote by $\mathbf{s}$ the syndrome vector corresponding to $X$ check operators after an error $E\in\{I,Z\}^n$ occurs.
The syndrome $\mathbf{s}$ satisfies $\mathbf{s}[i]=1$ if the $X$ check operator $S_i$ anti-commutes with $E$, and $\mathbf{s}[i]=0$ otherwise.
We refer to the set of $X$ check operators that anti-commute with $E$ as \textit{defects}.
We also define a binary noise vector $\mathbf{e}$, for which $\mathbf{e}[i]=1$ if an error $E\in\{I,Z\}^n$ acts non-trivially on qubit $i$, and $\mathbf{e}[i]=0$ otherwise.

The minimum-weight perfect matching (MWPM) decoder can be used when each single qubit $Z$ operator anti-commutes with two $X$ stabilisers.
When this property is satisfied, we can define a \textit{matching graph} $G$, where we associate each node of $G$ with an $X$ check operator, and each edge with a single qubit $Z$ error.
Each error $E\in\{I,Z\}^n$ then corresponds to a subset of edges called a \textit{1-chain}, and each syndrome corresponds to the subset of nodes associated with the defects.
Minimum weight decoding of $Z$ errors then corresponds to finding the smallest 1-chain that has the defect nodes at its boundary.
If the probability $p_i$ of each qubit $i$ suffering a $Z$ error is different, then we can assign each edge a weight $w_i=\log((1-p_i)/p_i)$~\cite{dennis2002topological}.
We see that the probability of an error $E$ occurring,
\begin{equation}
p(E)=\prod_i (1-p_i)^{(1-\mathbf{e}[i])} p_i^{\mathbf{e}[i]}=\prod_i (1-p_i)\prod_i \left(\frac{p_i}{1-p_i}\right)^{\mathbf{e}[i]},
\end{equation}
satisfies $\log(p(E))=\sum_i \log(1-p_i)-\sum_i w_i\mathbf{e}[i]$, and hence more probable errors will have lower weight.

As an example, the matching graph for the $X$ (site) check operators of a distance 10 planar surface code is shown in \Cref{fig:matching_graph}.
At the boundary, where single qubit $Z$ errors anti-commute with only a single check operator, we also add boundary nodes (denoted with hollow squares), and all boundary nodes are connected to each other by edges of weight zero.
\Cref{fig:error_graph} shows an example of an error (red edges) and the corresponding defects (blue stars).
The next step of the MWPM decoder after measuring the check operators is to construct a \textit{syndrome graph} $V$, which has a node for each defect in $\mathbf{s}$.
In the full implementation of the MWPM decoder $V$ contains an edge for each pair of nodes, forming a complete graph.
The weight of each edge $(u,v)$ in $V$ is given by the length of the shortest path between the corresponding check operators in the original matching graph $G$.
\Cref{fig:syndrome_graph} shows an example of a syndrome graph for the surface code.
The complexity of the Boost graph library implementation of Dijkstra's algorithm for finding the shortest paths between a source node and all other nodes is $O(N\log(N)+M)$ for a graph with $N$ vertices and $M$ edges~\cite{siek2002boost}.
To construct the syndrome graph, we must solve the single source shortest path problem in $G$ for each of the $|\mathbf{s}|$ defects.

The next step is to solve the minimum-weight perfect matching problem in $V$.
A \textit{matching} of a graph is a set of edges such that no two edges in the matching share a common vertex, and a \textit{perfect matching} is a matching that includes all vertices in the graph.
A minimum-weight perfect matching of a graph is a perfect matching that has minimum weight (where the weight of the matching is the sum of the weights of the edges).
This problem can be solved using the blossom algorithm~\cite{edmonds1965paths}, and efficient implementations are provided by the Lemon~\cite{dezsHo2011lemon} and Blossom V~\cite{kolmogorov2009blossom} C++ graph libraries.
The Lemon implementation of minimum-weight perfect matching has complexity $O(NM\log(N))$ for a graph with $N$ nodes and $M$ edges.
\Cref{fig:mwpm_graph} shows the minimum-weight perfect matching of the syndrome graph in \Cref{fig:syndrome_graph}.
For each edge $(u,v)$ in the matching, the minimum weight path from $u$ to $v$ in $G$ is then included in the correction that is output by the decoder (see \Cref{fig:correction_graph}).
We refer to this standard version of the MWPM decoder as \textit{exact matching}, and the solution output by the decoder is guaranteed to be the minimum-weight solution.
When syndrome measurements themselves are noisy, measurements are repeated $O(L)$ times and the \textit{difference syndrome} (parity of consecutive measurements) is used to decode over a 3D matching graph instead~\cite{dennis2002topological}.

For a syndrome $\mathbf{s}$ and a matching graph $G$ with $N$ vertices and $M$ edges, the construction of the syndrome graph using Dijkstra's algorithm has complexity $O(|\mathbf{s}|(N\log(N)+M))$.
The runtime of the blossom algorithm on the complete syndrome graph is then $O(|\mathbf{s}|^3\log(|\mathbf{s}|))$.
Since we can take $|\mathbf{s}|=O(N)$ (where $N\coloneqq |G|$) and $M=\beta N$ for some constant $\beta$ (assuming our code is LDPC), then the Dijkstra step has runtime $O(N^2\log(N))$ and the blossom step has complexity $O(N^3\log(N))$.
For the 2D matching graph of a distance $L$ surface code, the runtime of the decoder is therefore dominated by the blossom algorithm with overall running time $O(L^6\log(L))$.

Note that the MWPM decoder is not restricted to decoding Pauli errors in stabiliser codes. Given an $r\times n$ binary parity check matrix $H$ for which the weight of each column is two, and a syndrome vector $\mathbf{s}\in\mathbb{F}_2^r$, MWPM will find a noise vector $\mathbf{e}\in\mathbb{F}_2^n$ with minimum weight $\sum_i w_i\mathbf{e}[i]$ satisfying $H\mathbf{e}=\mathbf{s}$, where $w_i\in\mathbb{R}^+$ is a non-negative weight associated with the $i$th bit.

\section{Local matching}\label{sec:local_matching}

\begin{figure}
\centering
\includegraphics[trim={3cm 0 1.5cm 0},width=0.4\columnwidth]{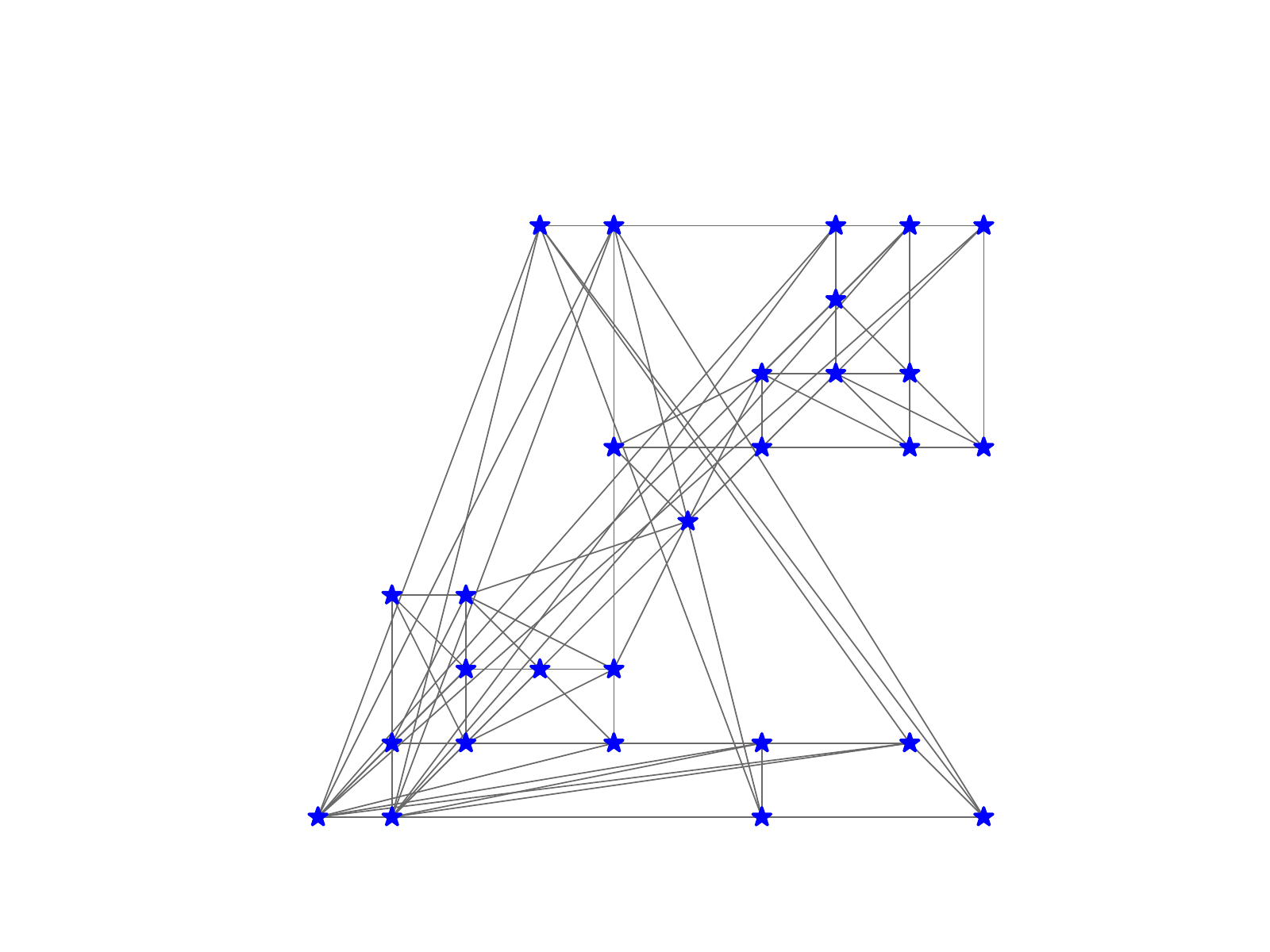}
\caption{The syndrome graph for the syndrome in \Cref{fig:mwpm_decoder} using the local matching decoder with $m=5$.}
\label{fig:local_matching_syndrome_graph}
\end{figure}

As well as exact matching, PyMatching includes a variant of the MWPM decoder which we call \textit{local matching}.
In local matching, first introduced in Ref.~\cite{higgott2020subsystem}, the syndrome graph is no longer chosen to be a complete graph of the defects, but instead contains only a subset of the edges, reducing the computational complexity of the algorithm.
The local matching decoder has a parameter $m$, which determines how sparse the syndrome graph is.
We denote the syndrome graph constructed for local matching with parameter $m$ by $V_m$.
We construct $V_m$ by adding an edge from each defect to its $m$ closest defects in $G$ (where here the distance $d(u,v)$ between two nodes $u$ and $v$ in $G$ is given by the length of the shortest path between them).
The weight of each edge $(u,v)$ in $V_m$ is again given by the distance $d(u,v)$ between them in $G$.

In order to construct $V_m$, we use a variant of Dijkstra's algorithm, which we call the local Dijkstra algorithm, to find the distance to each of the $m$ closest defects from defect $i$ in $G$.
The local Dijkstra algorithm is given in \Cref{alg:local_dijkstra}, and can also be found in Ref.~\cite{higgott2020subsystem}.
The difference with Dijkstra's algorithm is that the local Dijkstra algorithm keeps track of which defects in $G$ have been examined, and stops once $m$ defects have been examined (a vertex is examined when it is removed from the priority queue $Q$ with \textit{ExtractMin}, see \Cref{alg:local_dijkstra}).
We also track which elements of the distances $d$ and predecessors $p$ arrays have been updated in each use of the algorithm, such that only these elements need to be reset before the next use of the decoder (rather than having to reset each array in full in $O(|G|)$ time).
Provided $m$ is not too small, we find that $V_m$ is connected with high probability for typical use cases (for $m>10$ we empirically find that $V_m$ is almost guaranteed to be connected in 2D and 3D matching graphs).
However, it is still possible that $V_m$ can be disconnected, especially for small $m$, and if any of its connected components contains an odd number of defects then a matching cannot be found, even if the parity of the overall syndrome is even.
To prevent this rare problem from occurring, we check if $V_m$ is disconnected before running the blossom algorithm. 
If $V_m$ is disconnected then we increase $m$ by one and recompute the syndrome graph, repeating until the syndrome graph is connected.
In \Cref{fig:local_matching_syndrome_graph} we show the syndrome graph constructed by the local matching decoder for $m=5$, using the same syndrome as in \Cref{fig:syndrome_graph} where the complete syndrome graph used by full matching is shown.
Local matching is not guaranteed to find the minimum-weight solution for small $m$, however as shown in \Cref{sec:benchmarks} it is a very good approximation of exact matching while having reduced computational complexity.
Since the syndrome graph in local matching has $|\mathbf{s}|=O(N)$ vertices and $O(Nm)$ edges, the blossom step of the local matching decoder has complexity $O(N^2m\log(N))$.
The running time of the local Dijkstra step depends on how the defects are distributed in $G$.
Let us assume that the defects are uniformly distributed such that only $O(mN/|\mathbf{s}|)$ vertices in $G$ are examined by the local Dijkstra algorithm.
If we further assume that $G$ has bounded degree (as is the case for an LDPC code) and that $N/|\mathbf{s}|\in O(1)$ (assuming some fixed error rate), then the runtime of the local Dijkstra algorithm in this case is equivalent to running the standard Dijkstra algorithm on a bounded-degree graph with $O(m)$ vertices and $O(m)$ edges, with a running time of $O(m\log(m))$ for each source vertex, leading to an overall running time of $O(Nm\log(m))$ required to construct the syndrome graph.
For a 2D matching graph used to decode a distance $L$ surface code, the total running time, dominated by the blossom algorithm, is $O(L^4m\log(L))$.

\begin{algorithm}[h]
\Fn{LocalDijkstra($G$, $\mathbf{s}$, $m$, $i$)}{
For each $u\in G$, $d[u]=\infty$, $p[u]=u$\;
$d[i]=0$\;
Initialise priority queue $Q$\;
$Q.insert(i)$\;
Initialise empty list of found defects $l$\;
 \While{$Q$ is not empty and $length(l)<m$}{
  $u=Q.ExtractMin()$\Comment*{Examine vertex $u$}
  \If{$\mathbf{s}[u]=1$}{
  	$l.Insert(u)$\;
  }
  \For{each vertex $v$ adjacent to $u$ in $G$}{
  	\If{$weight(u,v)+d[u]<d[v]$}{
		$d[v]=weight(u,v)+d[u]$\;
		$p[v]=u$\;
		\eIf{$d[v]$ previously equal to $\infty$}{
			$Q.Insert(v)$\;
		}{
			$Q.DecreaseKey(v)$\;	
		}
	}
  }
  }
 }
  \caption{Local Dijkstra Algorithm}
 \label{alg:local_dijkstra}
\end{algorithm}

Similar strategies for reducing the computational complexity of the MWPM decoder have been considered before. 
The decoder proposed in Ref.~\cite{fowler2012towards} initially only connects vertices in the syndrome graph if they are geometrically close in the 3D lattice of a surface code matching graph.
While this approach was shown to be effective for the planar surface code, it is more challenging to generalise to hyperbolic codes~\cite{breuckmann2016constructions}, and the Euclidean distance between defects does not take into account the edge weights along the shortest path between them.
The approach in Ref.~\cite{xu2019high} still constructs a complete syndrome graph, but only calculates the exact edge weights using Dijkstra's algorithm for nodes within a chosen threshold distance $c$ of each other.
While this method reduces the complexity of calculating shortest paths with Dijkstra's algorithm, the complete syndrome graph is still input to the blossom algorithm, unlike our local matching decoder, for which the syndrome graph itself is sparse.
Furthermore, setting an appropriate threshold $c$ requires analysing the edge weights in the matching graph.
While these alternative approaches are effective in the specific contexts in which they are applied, our local matching decoder offers more flexibility, since it does not need to be tailored to the specific geometry of the quantum code, nor to the edge weights in the matching graph.

Further improvements in runtime can be had by using decoders that do not rely on the blossom algorithm.
Of particular note is the Union-Find decoder~\cite{delfosse2017almost}, which has a runtime almost linear in the number of nodes and, owing to its simplicity, is also amenable to fast implementations in hardware~\cite{das2020scalable}.
The Union-Find decoder typically has lower thresholds than MWPM, with a threshold of 9.9\% for the surface code, compared to 10.3\% using MWPM, although there has recently been progress developing variants of Union-Find that have improved thresholds~\cite{huang2020fault,meinerz2021scalable,hu2020quasilinear}.
Regardless of these developments, we expect that MWPM will remain a useful decoder for benchmarking the performance of decoders and quantum error correcting codes for the foreseeable future.

\section{Benchmarks}\label{sec:benchmarks}
\subsection{Performance}

\begin{figure}
\centering
     \begin{subfigure}[b]{0.45\textwidth}
         \centering
	\includegraphics[width=\columnwidth]{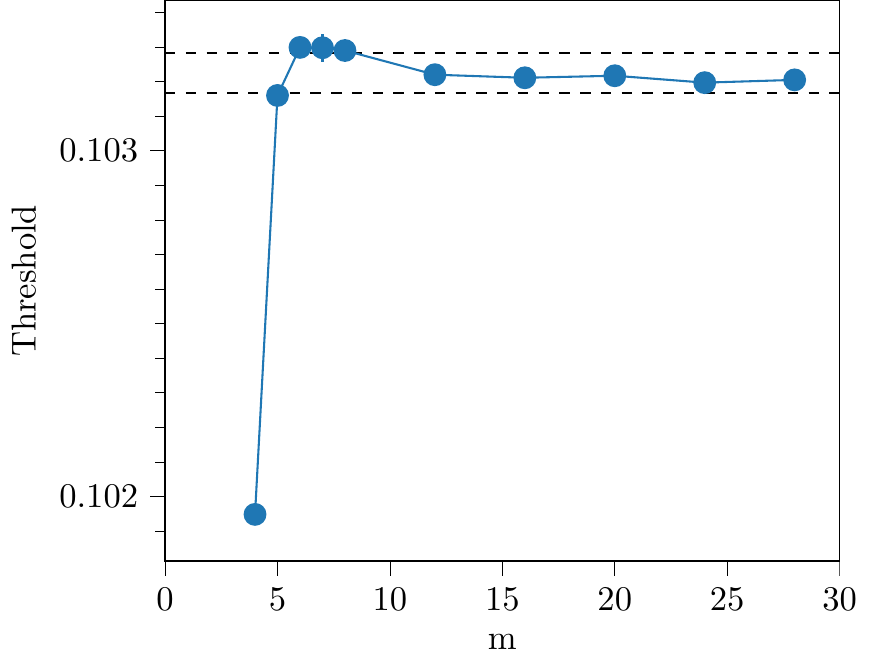}
	\caption{}
	\end{subfigure}
	\hfill
     \begin{subfigure}[b]{0.45\textwidth}
         \centering
	\includegraphics[width=\columnwidth]{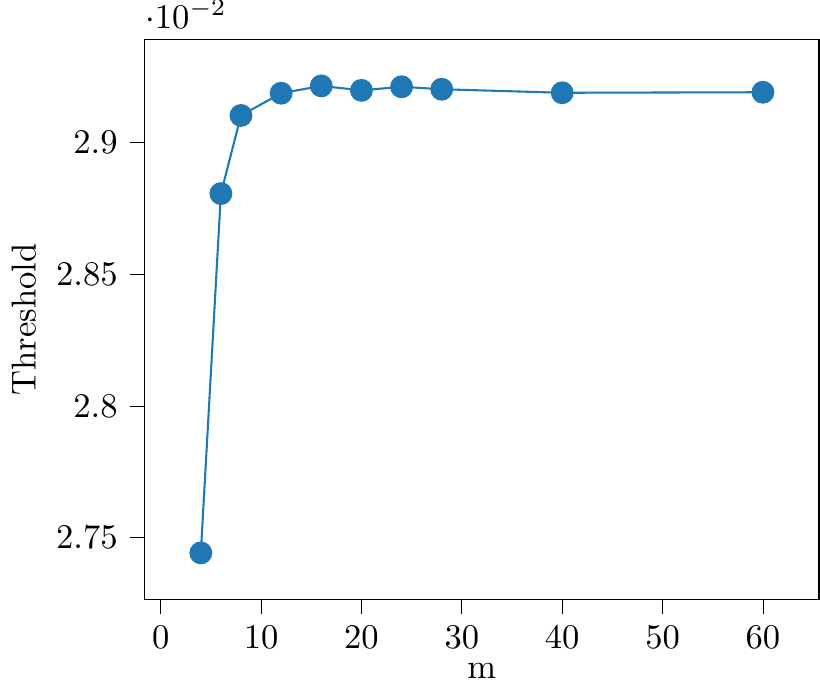}
	\caption{}
	\label{fig:thresholds_vs_m_phenom}
	\end{subfigure}
	\caption{Threshold of the local matching decoder as a function of $m$ for the 2D toric code, determined using lattice sizes $L=24,28,32,36$ and the critical exponent method of Ref.~\cite{wang2003confinement}. In (a) an independent noise model is used with perfect syndrome measurements, and the dashed lines show the $1\sigma$ lower and upper bounds on the threshold using full (exact) matching. In (b), a phenomenological noise model is used, for which both syndrome measurement and single qubit $Z$ errors occur with the same probability $p$. Syndrome measurements are repeated $L$ times, and decoding takes place over a 3D (2D + time) matching graph.}
	\label{fig:thresholds_vs_m}
\end{figure}

The parameterisation of the local matching decoder by the number of neighbours $m$ used to construct the syndrome graph allows for a trade-off between speed and the closeness of the decoder to exact matching.
By setting $m=|\mathbf{s}|-1$, where $|\mathbf{s}|$ is the Hamming weight of the syndrome, we recover exact matching at the expense of higher computational complexity.
By instead setting $m$ to a small constant, we get improved computational complexity and will still find a perfect matching, but the weight of the perfect matching is no longer guaranteed to be minimal.
However, as the benchmarks in this section show, the solutions output by local matching can still agree with exact matching with very high probability, even for small $m$.

\begin{figure}
\centering
     \begin{subfigure}[b]{0.45\textwidth}
         \centering
	\includegraphics[width=\columnwidth]{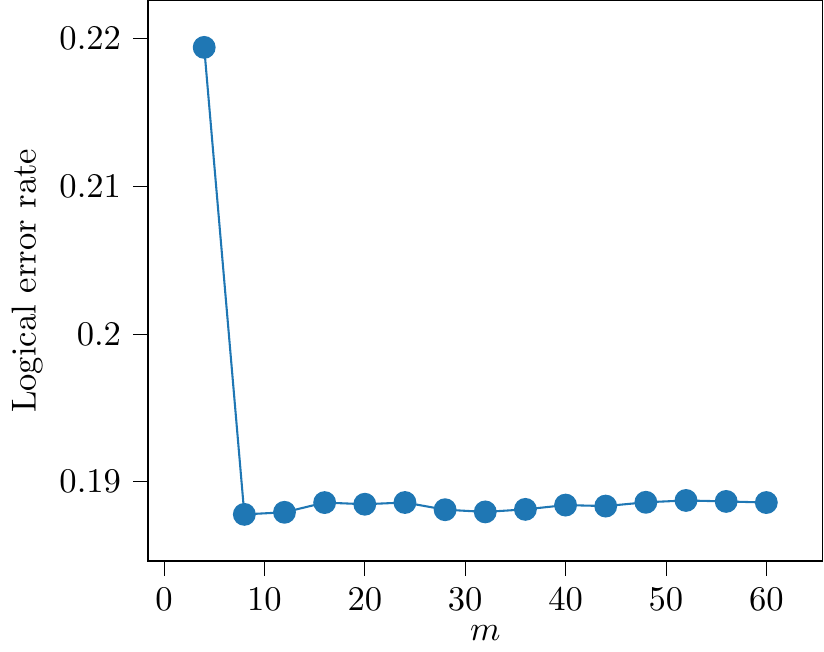}
	\caption{$p=0.1$, $L=60$, 2D}
	\label{fig:log_err_vs_m_2d}
	\end{subfigure}
	\hfill
     \begin{subfigure}[b]{0.45\textwidth}
         \centering
	\includegraphics[width=\columnwidth]{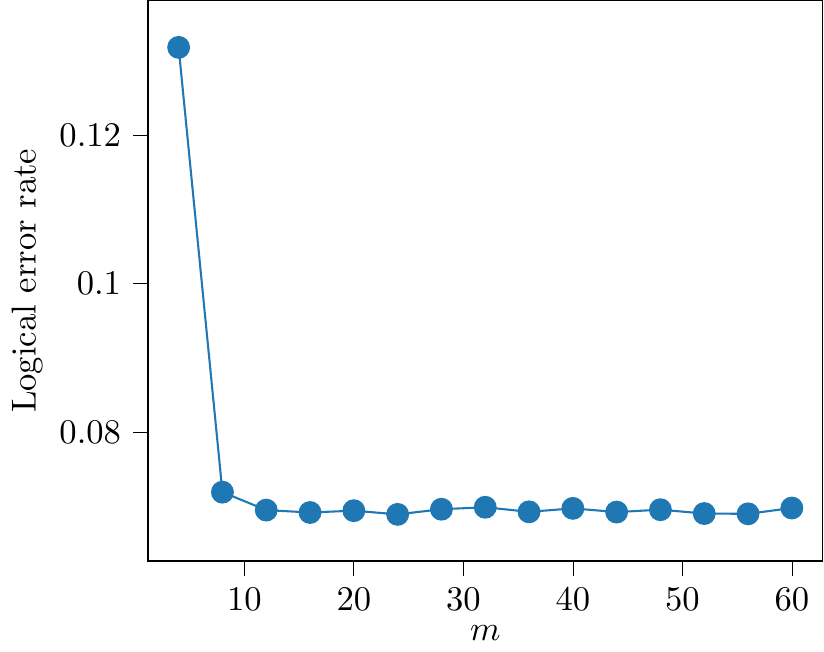}
	\caption{$p=0.029$, $L=20$, 3D}
	\label{fig:log_err_vs_m_3d}
	\end{subfigure}
	\caption{Logical error rate as a function of the number of neighbours $m$ used in the local matching decoder for the 2D toric code. (a) An $L=60$ toric code with $p=0.1$ and perfect syndrome measurements (2D matching graph). (b) An $L=20$ toric code with $p=0.029$ and noisy syndrome measurements (phenomenological noise model), with $L$ syndrome repeats (3D matching graph).}
	\label{fig:log_err_vs_m}
\end{figure}

In \Cref{fig:thresholds_vs_m} we show how the threshold of the 2D toric code varies as a function of $m$ for the local matching decoder.
All thresholds were estimated using lattice sizes $L=24,28,32,36$.
For perfect syndrome measurements, the threshold quickly converges to $0.10321(1)$ for $m\geq 12$, consistent with the expected value for exact matching~\cite{wang2003confinement}.
Interestingly, for $m=6,7,8$, the threshold of local matching is actually slightly higher than that of exact matching, and \Cref{fig:log_err_vs_m_2d} shows that the logical error rate for fixed $p$ near the threshold is also lower for $m=8$ than for exact matching.
This demonstrates that the approximation error from local matching for small $m$ does not necessarily degrade decoding performance, and can instead take advantage of degeneracy in the code to improve performance in some cases.
With noisy syndrome measurements (phenomenological noise model), local matching stabilises at a threshold of around 0.0292 for $m\geq 16$, also consistent with exact matching~\cite{wang2003confinement}.
In \Cref{fig:log_err_vs_m} the logical error rate is shown as a function of $m$, and we see that the logical error rate stabilises to a constant value for $m\geq 16$ for both perfect syndrome measurements and a phenomenological error model with a 3D matching graph.

\begin{figure}
\centering
\includegraphics[width=0.5\columnwidth]{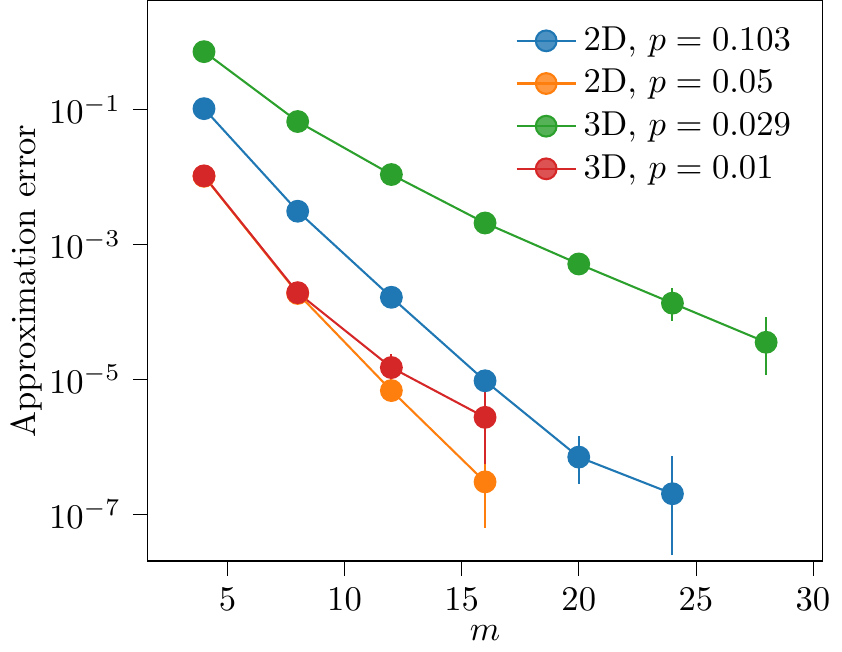}
\caption{Approximation error of the local matching decoder as a function of the number of neighbours $m$ for a $L=20$ toric code. The approximation error is defined as the fraction of runs for which the weight of the local matching differs from exact matching. Plots labelled 2D assume perfect syndrome measurements, whereas those labelled 3D assume $L=20$ rounds of noisy syndrome measurements (phenomenological noise model).}
\label{fig:approximation_error}
\end{figure}

To better understand how well local matching approximates exact matching, we also analyse the \textit{approximation error} of local matching, defined as the fraction of runs for which the weight of the local matching solution differs from that of exact matching.
From \Cref{fig:approximation_error} we see that the approximation error decreases exponentially with $m$, and also decreases with $p$.
With perfect syndrome measurements, the approximation error is less than $10^{-6}$ for $m\geq 20$.
With a phenomenological noise model near the threshold, the approximation error decreases more slowly and is still $\approx 10^{-3}$ at $m=20$.
However, the approximation error only provides an upper bound on the difference in logical error rate between local matching and exact matching, and as shown in \Cref{fig:thresholds_vs_m_phenom} and \Cref{fig:log_err_vs_m_3d}, we have not observed statistically significant differences in the logical error rate and threshold from varying $m$ for $m\geq 16$.

\subsection{Speed}

\begin{figure}
	\centering
     \begin{subfigure}[b]{0.45\textwidth}
         \centering
         \includegraphics[width=\columnwidth]{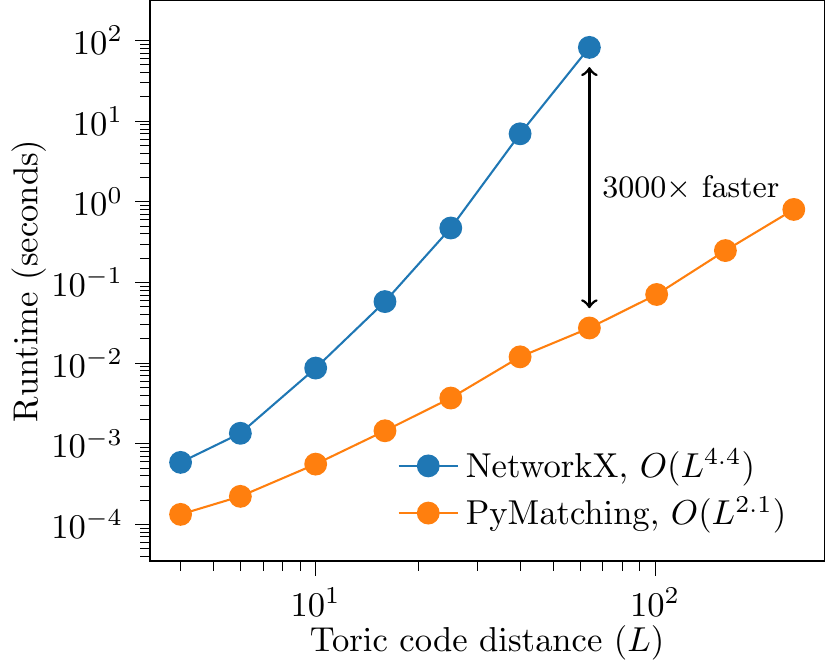}
         \caption{}
         \label{fig:pymatching_vs_networkx}
     \end{subfigure}
     \hfill
     \begin{subfigure}[b]{0.45\textwidth}
         \centering
         \includegraphics[width=\columnwidth]{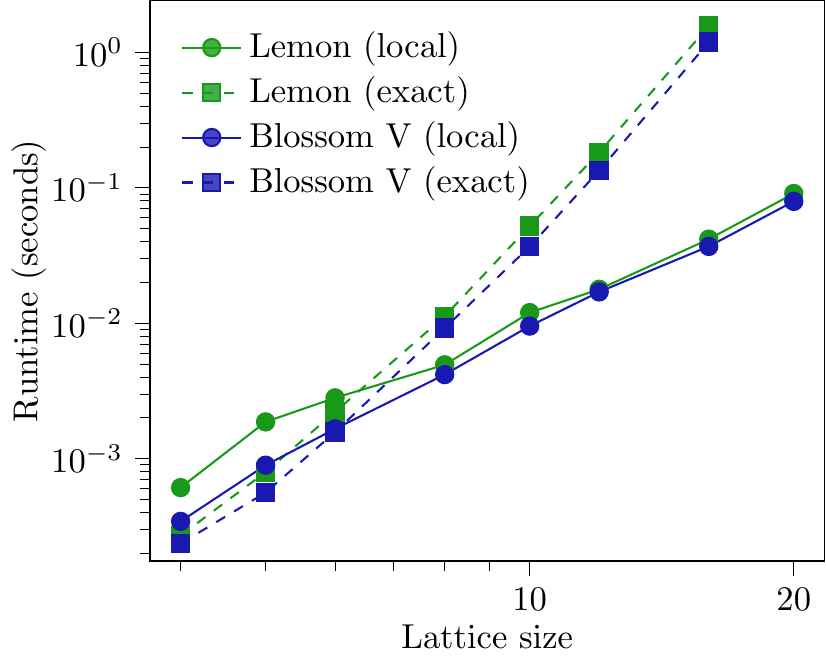}
         \caption{}
         \label{fig:lemon_vs_blossomv}
     \end{subfigure}
        \caption{Runtime of PyMatching for the toric code using a 2.8 GHz Intel Core i5 CPU. (a) The runtime of the PyMatching implementation of local matching for the toric code using $m=20$, with an independent noise model ($p=0.05$) and perfect syndrome measurements. For comparison, the runtime of an implementation of full matching using NetworkX is also shown. (b) Runtime comparisons of PyMatching for a 3D toric matching graph with a probability of $p=0.05$ of each edge being flipped. Local and full matching is shown with solid and dashed lines, respectfully. Results using the Lemon graph library (as used in the latest version of PyMatching) for the blossom algorithm are shown in green, and results using the Blossom V library are shown in blue.}
        \label{fig:pymatching_timing}
\end{figure}

As shown in \Cref{fig:pymatching_vs_networkx}, PyMatching can be several orders of magnitude faster than a NetworkX implementation of the minimum-weight perfect matching decoder.
This speedup is partly due to the fact that the core algorithms in PyMatching are implemented in C++ with the help of the excellent Lemon~\cite{dezsHo2011lemon} and Boost Graph libraries.
However, there is also a scaling advantage from the use of the local matching algorithm, rather than exact matching, leading to an empirically determined runtime scaling of around $O(L^{2.1})$ for $m=20$ (slightly worse than linear in the number of nodes), compared to $O(L^{4.4})$ for NetworkX.
These empirically-determined running times are considerably better than the expected worst case running time of $O(L^6\log(L))$ for exact matching and $O(L^4m\log(L))$ for local matching.
This suggests that the typical running time of the blossom algorithm is much better than its worst case complexity for matching graphs that typically arise in quantum error correction.
Indeed, for both local and exact matching, the empirically determined scaling is similar to the expected complexity of just the syndrome graph construction stage (Dijkstra or local Dijkstra) of the decoder.
The runtime scaling with $L$ we observe in local matching is also similar to the scaling observed for the variant of matching used in Ref.~\cite{fowler2012towardstiming}, although the implementation used in Ref.~\cite{fowler2012towardstiming} is specifically tailored to the surface code.

\begin{figure}
	\centering
     \includegraphics[width=0.4\columnwidth]{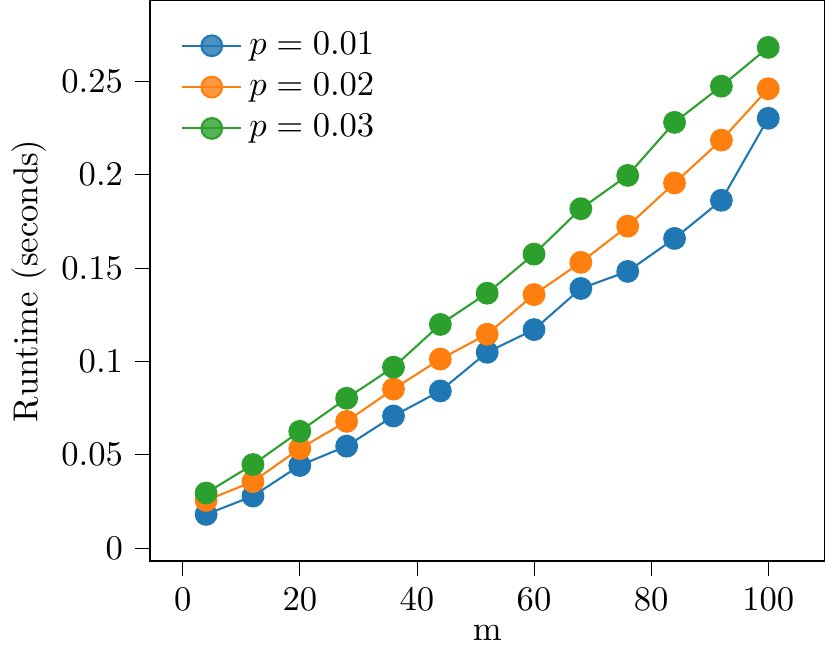}
        \caption{Runtime of PyMatching for different values of $m$ for the $L=20$ toric code, for three different error probabilities $p$. Pauli $Z$ errors and syndrome measurement errors both occur with probability $p$, and decoding takes place over a 3D matching graph (with $L$ syndrome measurement repeats).}
        \label{fig:pymatching_timing_vary_m}
\end{figure}

A commonly used C++ implementation of the blossom algorithm in the quantum error correction community  is the Blossom V library, which has excellent performance~\cite{kolmogorov2009blossom}.
However, Blossom V does not have a permissive software license and is therefore not used by PyMatching.
PyMatching instead uses the Lemon C++ library, which also has an efficient implementation of the Blossom algorithm but has a permissive, open-source license (the Boost license)~\cite{dezsHo2011lemon}.
In \Cref{fig:lemon_vs_blossomv}, we compare the performance of the Lemon and Blossom V C++ libraries when used by PyMatching for both exact and local ($m=20$) matching, for decoding a 3D toric matching graph with $p=0.05$.
For local matching at large $L$, the Blossom V library is only around 10-20\% faster than Lemon, and for exact matching Blossom V is around 20-30\% faster.
While this demonstrates that a small performance improvement can be achieved by using Blossom V instead of Lemon, we only use Lemon in PyMatching owing to its permissive, open-source license.
\Cref{fig:lemon_vs_blossomv} also demonstrates that, for the toric code and a phenomenological noise model, local matching is faster than exact matching for $L> 7$, and is more than one order of magnitude faster than exact matching for lattice sizes $L\geq 20$ often considered in fault-tolerance simulations.
For very small matching graphs, exact matching can be slightly faster, since PyMatching pre-computes the shortest paths between all pairs of nodes for exact matching but computes shortest paths on the fly for local matching.
In \Cref{fig:pymatching_timing_vary_m} we show how the runtime of PyMatching varies as a function of $m$.
We find that, for $m\geq 20$, the running time scales linearly with $m$ for the 2D toric code using a phenomenological noise model, both at and below threshold, which is consistent (up to logs) with the expected running time of $O(L^4m\log(L)+L^2m\log(m))$ found in \Cref{sec:local_matching}.
All timing analysis was done using a 2.8 GHz Intel Core i5 processor.

\section{Usage}

In this section, we will give some examples of how to use PyMatching, and we direct the reader to the documentation for more detailed examples.
While the algorithms for PyMatching are written in C++, all the core functionality is available via the Python bindings, which provide a simple interface.
The PyMatching Python package can be downloaded and installed from the Python Package Index with the command \pythoninline{pip install pymatching}.
The most simple way to input a quantum code to PyMatching is using a check matrix.
A check matrix $H$ is a binary matrix for which each element $H_{ij}$ is non-zero only if the $i$th check operator acts non-trivially on the $j$th qubit.
The first step is to construct a \pythoninline{pymatching.Matching} object, as shown here for a repetition code:
\begin{python}
    import numpy as np
    from pymatching import Matching
    
    H = np.array([
        [1,1,0,0,0],
        [0,1,1,0,0],
        [0,0,1,1,0],
        [0,0,0,1,1]
    ])
    m = Matching(H)
\end{python}
Note that the check matrix $H$ can also be supplied as a \pythoninline{scipy.sparse} matrix, which is more memory efficient for larger codes.
PyMatching can then be used to decode a binary syndrome $s$ using the \pythoninline{Matching.decode} method:
\begin{python}
    noise = np.array([0,0,1,1,0])
    s = H@noise %
    c = m.decode(s)
\end{python}
The correction $c$ is also a binary NumPy array, and the $i$th element of $c$ is nonzero only if the correction acts non-trivially on qubit $i$.

By default, PyMatching uses the local matching algorithm with $m=30$, however this can be changed by specifying the \pythoninline{num_neighbours} parameter when decoding. For example, to use $m=40$ in local matching we can use
\begin{python}
    c = m.decode(s, num_neighbours=40)
\end{python}
PyMatching can also be used to implement exact matching by instead setting \pythoninline{num_neighbours=None}.
When this option is used, the shortest paths between all pairs of nodes in the matching graph are computed and stored the first time \pythoninline{decode} is called, and then reused for later calls to \pythoninline{decode}.
While storing the shortest paths between all pairs of nodes speeds up exact matching, the memory requirement can be prohibitive for very large matching graphs.
Note that exact matching can also be used by setting \pythoninline{num_neighbours} to one less than the number of defects $|\mathbf{s}|-1$ (or any larger integer), since local matching is identical to exact matching in this limit.
While the memory requirements using this method are lower than when setting \pythoninline{num_neighbours=None}, the computational complexity is higher, since all distances between defects are computed on the fly.
It is generally recommended to stick to using local matching with \pythoninline{num_neighbours} set to a small constant, since it is faster than exact matching for all but the smallest matching graphs, and has modest memory requirements, while retaining almost identical decoding performance.

PyMatching can also be used to handle weighted edges, repeated noisy syndrome measurements, boundary nodes and hook errors (in which a single edge in the matching graph corresponds to more than one qubit error).
To help handle these use cases, PyMatching allows the \pythoninline{Matching} object to be constructed using a NetworkX graph, rather than a check matrix.
For example, using the quantum repetition code example again, we can construct the corresponding \pythoninline{Matching} object by first constructing a NetworkX graph:
\begin{python}
    import networkx as nx
    
    p = 0.2
    w = np.log((1-p)/p)
    g = nx.Graph()
    g.add_edge(0, 1, qubit_id=0, weight=w, error_probability=p)
    g.add_edge(1, 2, qubit_id=1, weight=w, error_probability=p)
    g.add_edge(2, 3, qubit_id=2, weight=w, error_probability=p)
    g.add_edge(3, 4, qubit_id=3, weight=w, error_probability=p)
    g.add_edge(4, 5, qubit_id=4, weight=w, error_probability=p)
\end{python}
where here each edge corresponds to a qubit that suffers an error with probability $p=0.2$ and is assigned a weight $\log(1-p)/p$.
With this approach, we are now able to add a hook error for which a single fault can lead to an error on more than one qubit.
This hook error corresponds to an edge in the matching graph for which the \pythoninline{qubit_id} is now a set of integers, and we will also give it a different error probability $p_2=0.12$ in this example:
\begin{python}
    p2 = 0.12
    w2 = np.log((1-p2)/p2)
    g.add_edge(2, 4, qubit_id={2, 3}, weight=w2, error_probability=p2)
\end{python}
Since nodes $0$ and $5$ are incident to only a single edge and do not correspond to stabilisers, we will specify that they are boundary nodes
\begin{python}
    g.node[0]['is_boundary'] = True
    g.node[5]['is_boundary'] = True
\end{python}
and then connect these boundary nodes with a boundary edge of weight zero that does not correspond to an error:
\begin{python}
    g.add_edge(0, 5, weight=0.0, qubit_id=-1, error_probability=0.0)
\end{python}
The \pythoninline{Matching} object corresponding to this graph can now be constructed with:
\begin{python}
    m = Matching(g)
\end{python}
If boundary nodes are specified and the syndrome supplied to \pythoninline{m.decode(s)} has odd parity, then one of the boundary nodes is flipped by PyMatching when decoding to ensure that the defects in the matching graph have even parity (otherwise a perfect matching does not exist).
If the optional \pythoninline{error_probability} attribute is specified for every edge in the graph, then PyMatching can also be used to simulate a stochastic noise model in which each edge is independently flipped with its corresponding \pythoninline{error_probability} using the \pythoninline{m.add_noise()} method.
Using all of this additional functionality, PyMatching can be used to simulate and decode circuit-level noise models in stabiliser measurement schedules, and was used for this purpose in Ref.~\cite{higgott2020subsystem}.

\section{Conclusions}

In this work, we have introduced PyMatching, a fast, open-source Python package for decoding quantum error correcting codes with the minimum-weight perfect matching (MWPM) decoder.
PyMatching uses a variant of the standard MWPM decoder, called local matching, which only allows nodes in the syndrome graph to be matched to its closest neighbours.
The benchmarks we have presented in this work demonstrate that the PyMatching implementation of local matching can be several orders of magnitude faster than implementations of exact MWPM using NetworkX or Blossom V~\cite{kolmogorov2009blossom} for large matching graphs, while retaining almost identical decoding performance.
While further optimisations may be possible by tailoring the decoder to specific codes, e.g.~by using Manhattan distances or translational symmetry in place of a Dijkstra search for a simple square lattice, PyMatching is instead designed to be flexible, capable of efficiently decoding any quantum error correcting code amenable to MWPM decoding.
We hope that this combination of speed and flexibility will lead to PyMatching being a valuable tool for quantum error correction researchers, saving programming time as well as computational resources.

\section{Acknowledgements}

The author is grateful to be supported by Unitary Fund and the Engineering and Physical Sciences Research Council [grant number EP/L015242/1]. The author also thanks Will Zeng, Nathan Shammah, Mike Vasmer and Craig Gidney for helpful discussions, as well as Nikolas Breuckmann and Dan Browne for feedback on the manuscript.
We acknowledge the use of the UCL Myriad High Performance Computing Facility (Myriad@UCL), and associated support services, in the completion of this work.

\bibliographystyle{unsrt}
\bibliography{references}

\end{document}